\documentclass{aastex701}
\usepackage{booktabs}
\usepackage{amssymb}
\usepackage{svg}
\usepackage{amsmath}
\newcommand{\degree}{^{\circ}}
\usepackage{xcolor}
\usepackage{makecell}

\begin{document}

\title{Ultraviolet Reflected-Light Spectroscopy of Modern Earth Analogs with Starshade Observations}

\author[orcid=0009-0005-9785-9325]{Zahra Ahmed}
\affiliation{Department of Aeronautics and Astronautics, Stanford University, Stanford, CA}
\email[show]{zjahmed@stanford.edu}  

\author[orcid=0000-0002-2005-4407]{Stuart Shaklan} 
\affiliation{Jet Propulsion Laboratory, California Institute of Technology, Pasadena, CA}
\email{email@email.com}

\author[orcid=0000-0002-1830-8260]{Mario Damiano}
\affiliation{Jet Propulsion Laboratory, California Institute of Technology, Pasadena, CA}
\email{email@email.com}

\author[0000-0002-4675-9069]{Armen Tokadjian}
\affiliation{Jet Propulsion Laboratory, California Institute of Technology, Pasadena, CA}
\email{email@email.com}

\author[0000-0002-2949-2163]{Edward W. Schwieterman}
\affiliation{Department of Earth and Planetary Sciences, University of California, Riverside, CA}
\email{email@email.com}

\author[0000-0002-6165-4986]{Simone D'Amico}
\affiliation{Department of Aeronautics and Astronautics, Stanford University, Stanford, CA}
\email{email@email.com}

\begin{abstract}
Characterizing rocky, habitable zone planets in reflected light is a key science objective for future direct imaging missions, and the ultraviolet (UV) is uniquely suited for supporting the characterization of Earth-like atmospheres. The Hartley-Huggins ozone band at $0.25 ~\mu\mathrm{m}$ remains detectable at low spectral resolution down to $0.1\%$ PAL $\mathrm{O}_2$, providing a sensitive proxy for oxygen-bearing chemistry even for Proterozoic-like atmospheres where the visible and near-infrared $\mathrm{O}_2$ features become inaccessible. This work uses an end-to-end framework for UV starshade spectroscopy of modern Earth analogs that integrates high-fidelity observation simulation, conservative post-processing, and Bayesian spectral retrieval to quantify ozone detection significance. The framework is applied across 108 simulated scenes spanning a range of planet phase angles, system inclinations, and exozodiacal dust densities. The results establish that robust ozone detection with UV-only, $R=7$ reflected light spectroscopy requires a photometric $\mathrm{SNR} \sim 30$. UV starshade observations with the Habitable Worlds Observatory achieve this threshold within 3-day integrations for solar-type systems at 10 parsecs with inclinations up to $60\degree$ at favorable phases ($\leq 90\degree$). Above this SNR threshold, ozone detections are robust to systematic bias from imperfect exozodi calibration within $\sim30\%$, and the median retrieved ozone abundance remains accurate to within one order of magnitude of the true value. 

\end{abstract}



\section{Introduction}
\label{sec:intro}

Future high-contrast, direct imaging missions such as the Habitable Worlds Observatory (HWO) will enable the spectroscopic characterization of terrestrial planets in reflected light, a key priority science area identified in the 2020 Astrophysics Decadal Survey \citep{national_academies_of_sciences_engineering_and_medicine_pathways_2023}. Characterizing Earth-like planets in the habitable zone of Sun-like stars across ultraviolet (UV), visible, and near-infrared (NIR) wavelengths is essential to both detect biosignatures and also place them within the broader environmental context \citep{krissansen-totton_wavelength_2025, 2026ASPC..542..279A}.
The UV is of particular importance for Earth-like atmospheres because ozone ($\text{O}_3$), a photochemical byproduct of oxygen ($\text{O}_2$), exhibits a strong absorption feature at $0.25 ~\mu \text{m}$ that can serve as a potential proxy for constraining oxygen abundances and is detectable with low spectral resolution. While $\text{O}_2$ can be detected in multiple bands for modern Earth atmospheres, its visible and NIR signatures diminish at low abundances. In contrast, the Hartley-Huggins ozone band is present down to 0.1$\%$ of present atmospheric $\text{O}_2$ levels (PAL $\text{O}_2$), representative of Earth's atmosphere during the Proterozoic Eon \citep{schwieterman_exoplanet_2018, reinhard_false_2017, Lisman_surveying_2019, schwieterman_importance_2018}. Under these conditions, UV ozone absorption may provide one of the most sensitive and observable indicators of oxygen-bearing chemistry.

Obtaining the reflected light spectrum of an Earth-like exoplanet requires achieving a contrast level of $\sim10^{-10}$ at the image of the planet. Significant progress has been made towards achieving this contrast in the visible and NIR wavelengths with coronagraphs such as the Roman Coronagraphic Instrument (CGI) \citep{Poberezhkiy_cgi_2025}. However, achieving similar UV performance with a coronagraph imposes stringent wavefront stability requirements that remain technologically challenging \citep{gorkom_performance_2025, gorkom_progress_2025}. Starshades offer an alternative architecture that is independent of the telescope and therefore are not subject to the same stability constraints. By blocking the starlight before it can enter the telescope, starshades achieve deep contrast, broad bandwidth, small inner working angle, and high throughput, resulting in a powerful tool for exoplanet characterization \citep{Cash2006}. They are also particularly well-suited to UV observations because the shorter wavelengths enable smaller starshade designs that are easier to manufacture and maneuver \citep{shaklan_starshade_2023}.

Prior retrieval studies of reflected-light spectra have demonstrated the feasibility of constraining atmospheric composition of Earth-like planets \citep{young_inferring_2024, tokadjian_detectability_2024, damiano_reflected_2023, damiano_reflected_2022, smith_detecting_2020, feng_characterizing_2018, lupu_developing_2016} as well as explored the required signal-to-noise ratios and wavelength coverage needed for meaningful constraints \citep{salvador_earth_2025, latouf_bayesian_2024, latouf_bayesian_2023}. 
These studies typically adopt a background-limited noise model where photon noise from the astrophysical background dominates over the planet signal but where the systematic noise from this background and the instrument is assumed to be cleanly subtracted to yield a spectrum with well-characterized uncertainties.
For future direct imaging missions, however, the photon noise-limited assumption may not hold. Specifically, exozodiacal debris is likely to be a significant source of noise for future direct imaging observations of habitable zone planets that must be subtracted out during post-processing. There is significant uncertainty in the density and structure of the exozodiacal debris disks that missions such as HWO will image, making complete removal of systematic bias difficult \citep{currie_mitigating_2023}. The systematic residuals resulting from imperfect subtraction will depend on observing geometry, the fidelity of the exozodi model, and the post-processing pipeline itself. 

The Starshade Exoplanet Data Challenge \citep{Hu_et_al_2021, Damiano_et_al_2024} examined the impact of systematic exozodiacal disks and their structures on the detection of exoplanets of various sizes in a range of orbits. The exozodiacal backgrounds were estimated to 10\% with non-parametric independent component analysis. The study concluded that the residual background led to substantial false positives and negatives, emphasizing the importance of background calibration when estimating the science return of exoplanet missions.
Despite this, the impact of systematic backgrounds on retrieved abundances remains largely unexplored. 
Recent work by \cite{currie_exozodi_2026} addresses this gap for visible and NIR spectroscopy, finding that reliable abundance retrieval requires residual exozodi flux below $10^{-3}$ of the planet flux or $10^{-2}$ to preserve absorption band depth; however it is uncertain whether comparably strict requirements apply to UV-only ozone detection, given the depth of the Hartley-Huggins feature.
While precursor observations would be ideal to constrain exozodi levels and morphologies for individual targets \citep{stapelfeldt_nasa_2025}, it is important to understand how imperfect subtraction propagates into retrieval accuracy in the NUV.

To address this gap, this work extends an observation simulation and post-processing framework introduced in \cite{ahmed_ultraviolet_2025} to include Bayesian spectral retrieval.
Thus, unlike prior retrieval studies that assume a spectrum with idealized noise properties, this work is novel in its end-to-end modeling that connects realistic observation simulations and post-processing directly to atmospheric characterization, allowing systematics from imperfect background subtraction to propagate into the retrievals.
This framework is applied to UV starshade observations of modern Earth analogs, focusing on retrieving ozone abundances across a range of phase angles, exozodi levels, and system inclinations. 
By mapping where ozone can be robustly detected in the presence of systematic errors, this work (1) quantifies the science return of low-resolution UV starshade spectroscopy under realistic observing conditions and (2) assesses the impact of systematics on retrieval accuracy and precision.

The remainder of the paper proceeds as follows: Section~\ref{sec:sim_data} summarizes the observation simulation framework and parameters that are used to generate and post-process the UV starshade observations and ultimately obtain the noisy spectra. Section~\ref{sec:retrieval_setup} describes the retrieval set-up used to constrain the modern Earth atmosphere parameters and Section~\ref{sec:bayes_factor} describes the statistical analysis that is used to determine the significance of the ozone detection. 
Section~\ref{sec:results} describes the overall results of the study and Section~\ref{sec:discussion} discusses the specific trends of the retrievals over the parameter space and how they relate to the objectives of this work.
Finally, Section~\ref{sec:conclusion} provides conclusions and recommendations for future direct-imaging missions.

\begin{figure}[t!]
    \centering
    \includegraphics[width=\linewidth]{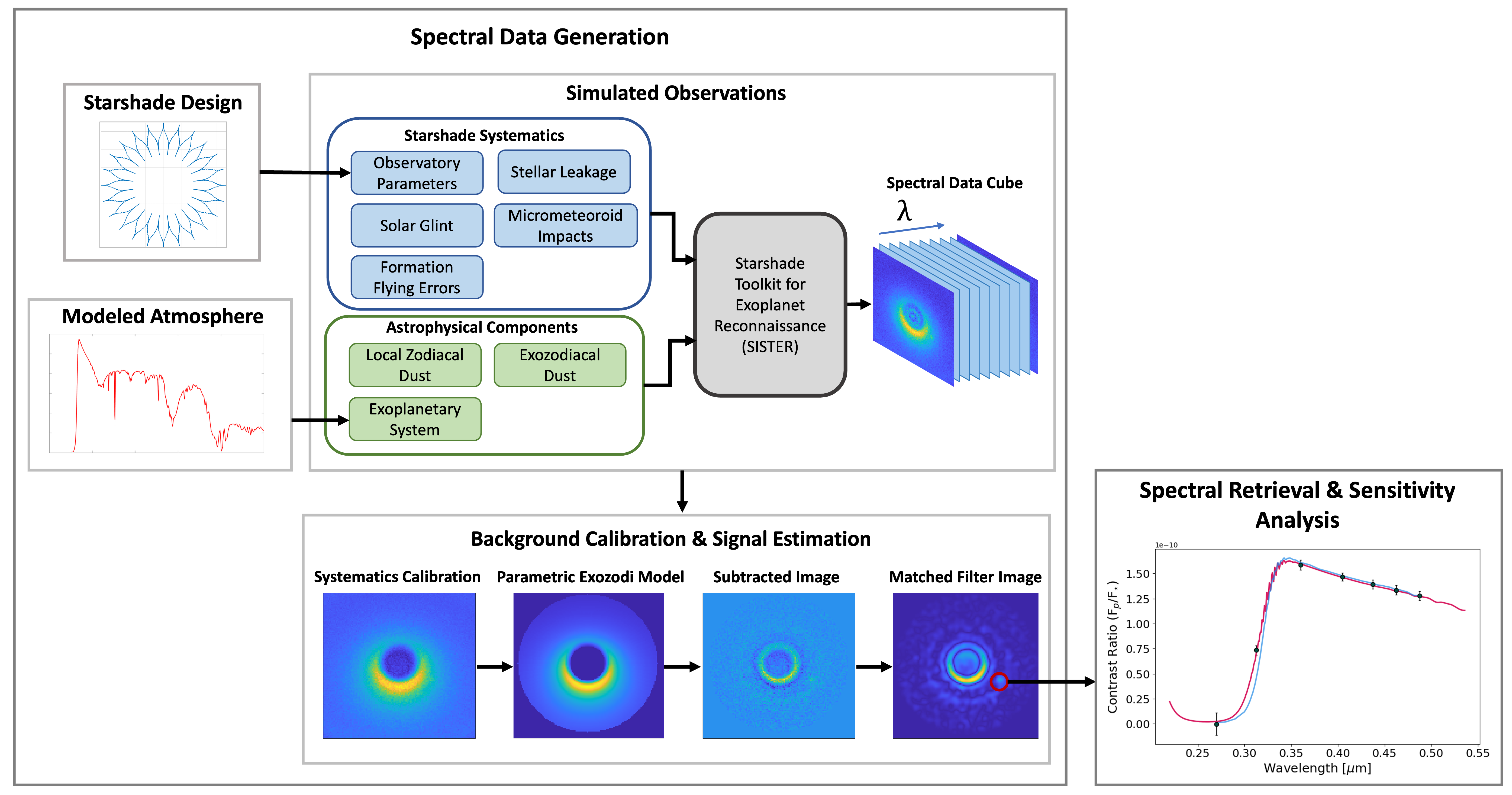}
    \caption{Overview of the end-to-end modeling and retrieval framework used in this work. The starshade design (Section~\ref{sec:starshade}) and simulated modern Earth atmosphere (Section~\ref{sec:sim_atmosphere}) are used as inputs to the simulated observations (Section~\ref{sec:systematics}), which are generated over a range of system inclinations, phase angles, and exozodiacal dust density. The resulting observations are post-processed to generate the spectral data (Section~\ref{sec:postprocessing}) that are used for the spectral retrievals and sensitivity analysis (Sections~\ref{sec:retrieval_setup}, \ref{sec:bayes_factor}).
    }
    \label{fig:paper_overview}
\end{figure}

\section{Spectral Data Generation}
\label{sec:sim_data}

The ultraviolet spectra analyzed in this work are generated using an observation simulation and post-processing framework developed in \cite{ahmed_ultraviolet_2025} and illustrated in Figure \ref{fig:paper_overview}. This framework is used to generate a suite of high-fidelity simulations that model starshade-specific instrument systematics, astrophysical noise from local and exozodiacal dust, an exoplanetary system, and photon and detector noise. The observatory, starshade, and exoplanet system parameters can be specified and tailored to any use case. In this work, the simulations are injected with a single modern Earth-like planet orbiting a solar-type star at 10 pc and assume 3-day exposure times. The input modern Earth albedo spectra is generated using the same forward model that is used for the retrievals, as will be discussed in Section~\ref{sec:sim_atmosphere}. The simulations are generated as three-dimensional data cubes with two spatial dimensions and one spectral dimension. 

The data cubes are processed through a pipeline that conservatively estimates the instrument systematics, fits and subtracts a parametric model of the exozodiacal disk, and finally applies a matched filter to maximize the signal-to-noise ratio (SNR) of the planet signal. Following post-processing, the planetary spectrum is extracted by sampling the filtered data cube at the planet location across each wavelength band. This process is repeated for 1000 Monte Carlo trials where photon noise, detector noise, and the calibration factor for the starshade systematics are independently sampled for each trial. The mean values across all trials are adopted as the measured spectrum, while the standard deviation at each wavelength defines the corresponding uncertainty.

This approach is designed to produce spectra that are representative of realistic starshade observations. As a result, the spectra are not purely photon noise-limited and instead provide conservative, noisy inputs for atmospheric retrieval analyses. For the purposes of determining the conditions under which ozone is detectable, this work varies the simulations over a range of system inclinations, planetary phase angles, and exozodiacal dust levels, totaling 108 unique simulated scenes and thereby enabling a sensitivity study of how these parameters affect spectral quality and retrieval performance. An overview of the simulated data and post-processing pipeline is presented in the following sections; full details are provided in \cite{ahmed_ultraviolet_2025}.

\subsection{Starshade-Observatory Overview}
\label{sec:starshade}

The observations are generated using the Starshade Imaging Simulation Toolkit for Exoplanet Reconnaissance (SISTER) \citep{hildebrandt_starshade_2021}. The simulations assume a 6-m diameter telescope, diffraction-limited at 250 nm, operating at a separation of 55,000 km with a 35~m diameter starshade. The telescope's diffraction-limited performance in the NUV allows it to resolve out exozodiacal light and perform observations that would be prohibitively long in the visible and NIR (\cite{mennesson_2024}). The starshade design was introduced in \cite{shaklan_starshade_2023} and is optimized for the $250-500$~nm near-UV bandpass with an inner working angle at the petal tips of 65 milli-arcseconds (mas). The nominal design achieves better than $5 \times 10^{-12}$ contrast at the petal tips over the full bandpass. \cite{Damiano_starshade_obs} have shown that this design is capable of up to 180 Earth-like planet observations in a 5 year mission without refueling. The individual imaging bands were updated from those used in \cite{ahmed_ultraviolet_2025} based on unpublished analysis\footnote[1]{Private communication with Bertrand Mennesson and Amber Young.} on the optimal UV bandpasses at low spectral resolution. This study determined that the $250-290$, $290-335$, and $335-385$~nm bands were optimal for distinguishing ozone from possible degenerate solutions for compounds such as sulfur dioxide ($\mathrm{SO}_2$) during retrievals for modern Earth atmospheres. The remaining imaging bands were selected to be consistent with the 25~nm wide bands used in ~\cite{ahmed_ultraviolet_2025}. Thus, the spectral resolution is approximately $R = 7$ at 250 nm, consistent with previous UV retrieval studies \citep{damiano_reflected_2023} and the LUVOIR \citep{roberge_large_2018} and HabEx \citep{gaudi_habitable_2020} mission studies. A summary of the observatory and starshade parameters is presented in Table~\ref{tab:obs_ss_params}.

\begin{table}[t]
    \centering
    \caption{Observatory and starshade parameters used in this work. Recreated from \cite{ahmed_ultraviolet_2025}.}
    \label{tab:obs_ss_params}
    \begin{tabular}{l l}
        \toprule       
        \textbf{Parameter} & \textbf{Value} \\
        \midrule
        \multicolumn{2}{l}{\textbf{Observatory Parameters}} \\
        \hline
        \addlinespace[0.5ex]

        Telescope Diameter & 6 m \\
        Optical Throughput & 0.35 \\
        Quantum Efficiency & 0.5 \\
        Frame Exposure Duration & 600 s \\
        Imaging Bandpass & 250--290, 290--335, 335--385, 385--425, \\
                         & 425--450, 450--475, 475--500~nm \\
        Pixel Scale & 3 mas/pixel \\
        Detector Gain & 1 \\
        Read Noise & 0.1 e$^-$/pixel/frame \\
        Dark Current & 0.0001 e$^-$/pixel/s \\
        Clock Induced Charge & 0.0013 e$^-$/pixel/frame \\

        \addlinespace[1ex]
        \hline
        \multicolumn{2}{l}{\textbf{Starshade Parameters}} \\
        \hline
        \addlinespace[0.5ex]

        Tip-to-Tip Diameter & 35 m \\
        Inner Working Angle (IWA) at Petal Tips & 65 mas \\
        Operating Bandpass & 250--500 nm \\
        Telescope--Starshade Separation & 55{,}000 km \\

        \bottomrule
    \end{tabular}
\end{table}

\subsection{Simulated Modern Earth Atmosphere}
\label{sec:sim_atmosphere}

\begin{figure}[tb]
    \centering
    \includegraphics[width=0.5\linewidth]{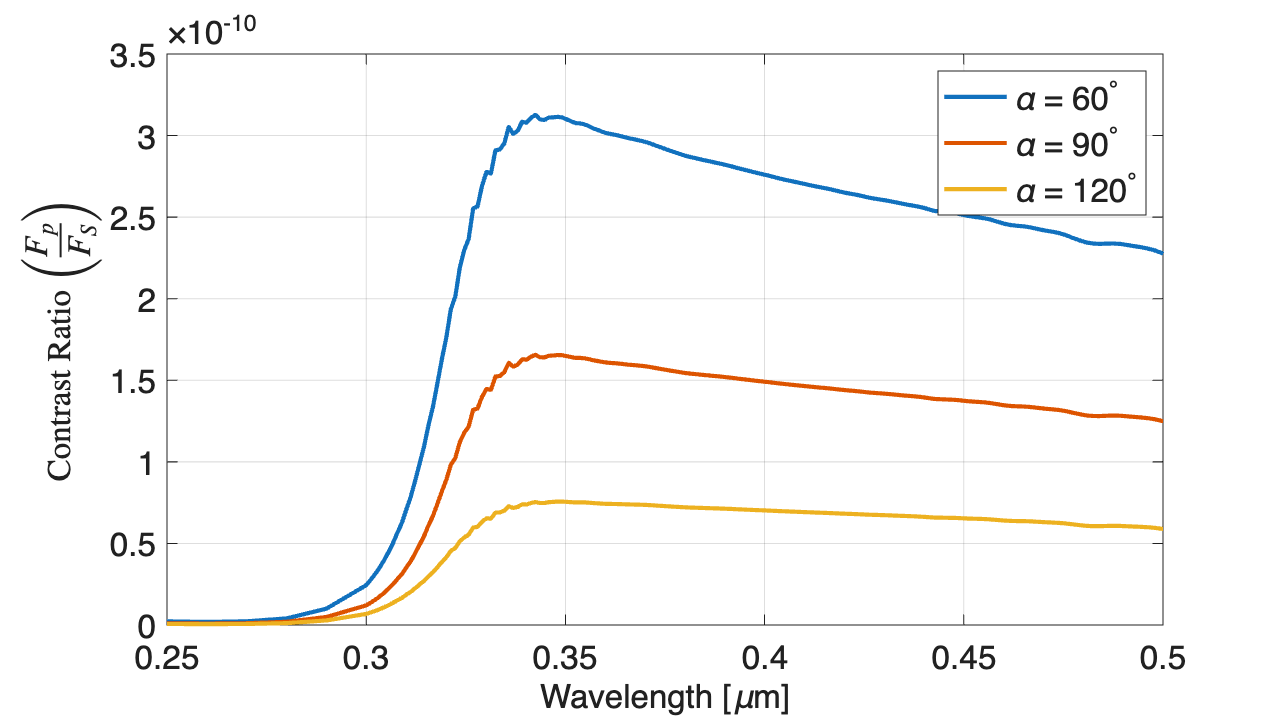}
    \caption{Modern Earth spectrum generated using ExoReL as a function of phase angle. The planet flux ratio scales with the phase angle, changing the depth of the 250 nm ozone feature.}
    \label{fig:modernEarthSpectrum}
\end{figure}

The simulated system includes a single Earth-like planet orbiting a solar-type star in a 1 AU circular orbit at 10 parsecs. The planet is injected at various angular positions in its orbit, referred to as the planet longitude. If the planetary system was observed edge-on, a longitude of $0^{\circ}$ corresponds to when the planet is directly behind the star, $90^{\circ}$ is at quadrature, and $180^{\circ}$ is when the planet is directly in front of the star. The system inclination is also varied from $0^\circ$ (face-on) to $90^\circ$ (edge-on) in $30^\circ$ increments. The planet longitude is independent from the system inclination, as it is defined in the orbital plane. The planet longitude and system inclination together determine the planetary phase angle $\alpha$, 
\begin{equation}
\label{eq:phase_angle}
    \alpha = cos^{-1}\left(cos\left( \beta \right)sin\left(i\right)\right)
\end{equation}
where $\beta$ is the planet longitude and $i$ is the system inclination. The planet phase angle describes what fraction of the planet is illuminated by the host star as viewed by a distant observer. Figure~\ref{fig:modernEarthSpectrum} shows how the spectrum scales as a function of phase angle. Smaller phase angles are more favorable as they maximize the observed planet-star flux ratio and observed depth of molecular features.
The modern Earth spectra shown in Figure~\ref{fig:modernEarthSpectrum} are used as the input to the simulated observations and were generated using the forward model from $\text{EXOREL}^{\mathcal{R}}$ \citep{damiano_exorelmathfrakr_2020} for a 1 $M_\oplus$, 1 $R_\oplus$ planet with the input values shown in Table~\ref{tab:modern_earth}. The same forward model is used for the atmospheric retrievals described in Section~\ref{sec:retrieval_setup}. The noiseless spectrum and corresponding orbital geometry parameters are used as inputs to the starshade simulations.

\begin{table}[tb]
    \centering
    \caption{Input parameters used to simulate a modern Earth atmosphere using $\text{EXOREL}^{\mathcal{R}}$. Parameters with an asterisk (*) are derived from other free parameters. The molecular abundances are specified as the log of the volume mixing ratios (VMR).}
    \begin{tabular}{ccc}
    \toprule
        \textbf{Parameter} & \textbf{Description} & \textbf{Input Value} \\
        \midrule
        $Log(P_0)^*$ & Surface Pressure (Pa) & 5.00  \\
        $Log(P_{top, H_2O})$ & Cloud Top (Pa)  & 4.85  \\
        $Log(D_{top, H_2O})$ & Cloud Depth (Pa)  & 4.0 \\
         -- & Cloud Fraction & 0.5 \\
        $A_g$ & Surface Albedo & 0.05 \\
        $\mu$ & Mean Molecular Weight & 28.827 \\
        $Log(CR_{H_2O})$ & Condensation Ratio & -3.5 \\
        $Log(\text{VMR}_{H_2O})$ & $H_2O$ & -2.523 \\
        $Log(\text{VMR}_{CH_4})$ & $CH_4$ & -5.70 \\
        $Log(\text{VMR}_{CO_2})$ & $CO_2$ & -3.40 \\
        $Log(\text{VMR}_{O_2})$ & $O_2$ &  -0.677 \\
        $Log(\text{VMR}_{O_3})$ & $O_3$ &  -6.155\\
        $Log(\text{VMR}_{N_2})^*$ & $N_2$ & -0.104\\     
    \bottomrule
    \end{tabular}
    \label{tab:modern_earth}
\end{table}

\subsection{Instrument and Astrophysical Systematics}
\label{sec:systematics}

In addition to the planet-star system, the simulated observations incorporate both instrument-specific systematics and astrophysical backgrounds. The starshade instrument systematics include stellar leakage from petal shape and position imperfections \citep{shaklan_starshade_2023}, solar glint from petal edges with conservative estimates of anti-reflection coating performance \citep{mckeithen_edges} and particulate contamination \citep{mckeithen_particulate}, sunlight and starlight leakage through micrometeoroid holes accumulated over a 10-year mission \citep{shaklan_2024}, multiple-scatter of sunlight from non-planar petals \citep{martin_scatter}, and stellar leakage from a 1-m lateral formation flying error. All of these terms are validated against prior laboratory experiments and analyses as detailed in \cite{ahmed_ultraviolet_2025}.

The astrophysical backgrounds consist of local zodiacal dust from the Solar System and exozodiacal debris disks, or exozodi, around the target system. The local zodiacal light is modeled as a spatially uniform background that is blocked by the starshade. The exozodiacal dust represents one of the dominant sources of noise for planet signal recovery with direct imaging. Rather than using smooth, axisymmetric disk models, this work employs scattered light images of resonant exozodiacal debris disks from \cite{currie_mitigating_2023}, which were generated using n-body simulations \citep{stark_transit_2011} that capture the asymmetric structures arising from planet-disk gravitational interactions. These simulations account for Poynting-Robertson drag, radiation pressure, and dust particle collisions, producing disks with mean motion resonant clumps that lead and trail the planet's orbital position. Such resonant structures create a non-uniform background around the planet that is more difficult to calibrate than a smooth disk, particularly for inclined systems where forward scattering amplifies the exozodi brightness in the region near the planet. The exozodi models span inclinations of $0^\circ$ (face-on), $30^\circ$, $60^\circ$, and $90^\circ$ (edge-on) and dust densities of 1, 5, 10, 20, 50, and 100 zodis, where 1 zodi corresponds to the dust density of the Solar System. These resonant disk models represent a conservative, worst-case scenario for exozodi calibration, making them well-suited for assessing the robustness of the spectral retrievals presented in this work.

The planetary system, instrument systematics, and zodi are incorporated into the simulated scene and convolved with the optical response of the telescope and starshade to produce the composite observations. Finally, photon and detector noise are added for a given integration time using SISTER's noise model. This includes photon noise from all scene components, dark current, read noise, and clock-induced charge, following the respective parameter values in Table~\ref{tab:obs_ss_params}.

For each simulation, the photometric SNR is computed in the 335-385nm bandpass directly from the planet and background cubes assuming a 3-day integration time as 

\begin{equation}
    \mathrm{SNR} = \dfrac{C_{p}}{\sqrt{C_{p} + C_{b} + D_C + R}}
\end{equation}

where $C_{p}$ and $C_{b}$ are the total planet and background counts, respectively, in a $1.4 ~\lambda/D$ aperture, $D_C$ is the dark current multiplied by the number of pixels in the aperture, and $R$ is the read noise multiplied by the number of pixels in the aperture and divided by the frame exposure duration. The photometric SNR therefore represents the background-limited floor for a given observing geometry and is independent of the post-processing fidelity.

\subsection{Background Calibration and Signal Estimation}
\label{sec:postprocessing}

The background calibration and signal estimation framework assumes that the planet location is known a priori. This assumption is consistent with the concept of operations for 
missions that are designed for detailed spectral characterization of previously detected planet candidates.

\begin{figure}
    \centering
    \includegraphics[width=0.95\linewidth]{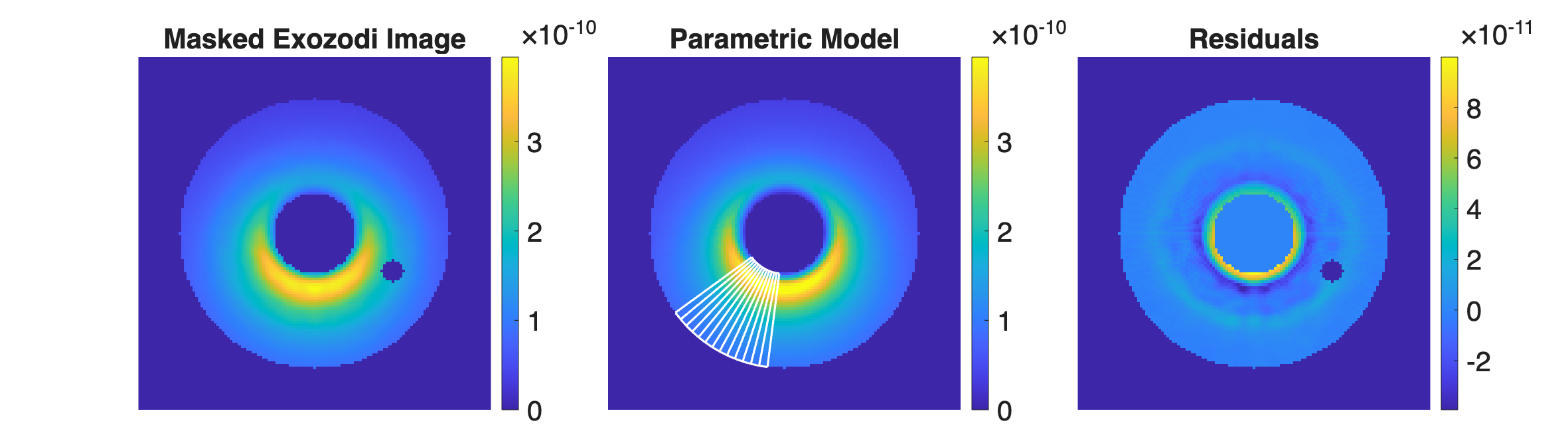}
    \caption{The exozodi estimation and subtraction process shown for a noiseless image of a $30\degree$ inclined system with 20 zodis and a planet injected at $\beta = 120\degree$. The images are shown for the 335-385 nm band. (\emph{Left}) The image is restricted to an annulus between 45 mas and 150 mas and the planet is masked out. (\emph{Middle}) The parametric model is constructed by fitting a second-order reciprocal power function to narrow azimuthal sectors. A subset of the sectors are shown in white overlaid over the resulting exozodi model. (\emph{Right}) The resonant ring structure is seen in the residuals after subtracting the model from the input image.}
    \label{fig:exozodi_postProce}
\end{figure}

After generating the simulated observations, the resulting data cubes are processed using a three-step pipeline. First, instrument systematics are imperfectly subtracted using a calibration factor drawn from a normal distribution centered on 1 (i.e. perfect calibration) and a standard deviation of 0.1 to account for realistic over- and under-subtraction. Second, the exozodiacal light is estimated and removed using a smooth parametric model. The parametric model is constructed by fitting a second-order reciprocal power function to the calibrated image, weighted by the starshade's normalized core throughput. This formulation is motivated by the expectation that scattered light from exozodiacal debris can be approximated by the inverse-square law, modulated by the radial transmission of the starshade. The coefficients for the parametric model are calculated using least squares for narrow azimuthal sectors of width $\Delta\theta = \pi / 50$. The model is constrained to an annulus with an inner radius of 45 mas and an outer radius of 150 mas, excluding the region within the starshade IWA, where transmission falls below $10\%$, and the region well outside the habitable zone. Before fitting, the planet is masked out using a 1.4 $\lambda$/D diameter aperture so that its flux does not bias the disk estimate. To suppress discontinuities introduced by the discrete azimuthal binning, a moving-average filter with a window of $\pi/10$ is applied to the fitted coefficients, and the resulting smooth model is subtracted from the calibrated image. This process is repeated for each wavelength band. Figure~\ref{fig:exozodi_postProce} illustrates the exozodi estimation and subtraction step for a noiseless image of an exozodi debris disk with 20 zodis and a planet injected at a longitude of $\beta = 120\degree$. Finally, after the exozodi is estimated out of the data cube, the residual image is convolved with a normalized off-axis PSF template (i.e., matched filtered) to maximize the signal-to-noise ratio \citep{ruffio_improving_2017}. Following these steps, the planet spectrum is extracted by measuring the recovered planet flux in each spectral channel.

In practice, the post-processing does not fully remove all background terms, and residual systematics remain in the data. The instrument calibration step allows for realistic over- and under-subtraction, and the smooth parametric exozodi model does not capture fine-scale resonant structure, as can be seen in the right panel of Figure~\ref{fig:exozodi_postProce}. Because the simulated exozodi disks include resonant clumps and azimuthal asymmetries, these features are not completely removed and persist as structured residuals in the final image. To quantify the impact of these residuals on retrieval performance, the mean bias is defined as the fractional difference between the extracted planet flux and the true input planet flux, averaged from the measurements between 335-500nm across all Monte Carlo trials. This measurement provides an estimate of the over-subtraction or under-subtraction in the continuum as a means to quantify the effectiveness of the post-processing pipeline.

\section{Retrieval Set-Up}
\label{sec:retrieval_setup}

The atmospheric retrievals are executed using the Exoplanetary Reflected Light Retrieval, $\text{EXOREL}^{\mathcal{R}}$, framework introduced in \citet{damiano_exorelmathfrakr_2020}. $\text{EXOREL}^{\mathcal{R}}$ has been used in previous studies of reflected light atmospheric characterization of small exoplanets \citep{damiano_reflected_2021, damiano_reflected_2022} including in the UV \citep{damiano_reflected_2023}. 

The full list of free parameters are tailored to the NUV analysis specified for this work and are reported in Table~\ref{tab:free_params}. The gas abundances are retrieved as partial pressures and therefore the surface pressure, $P_0$, is a derived quantity instead of a free parameter. 
Planetary mass and surface gravity are fixed in the retrievals due to their degeneracy with radius and surface albedo in reflected light.
The planet phase angle is also set as a fixed parameter because this work investigates how planetary phase angle affects the ability to retrieve molecular abundances and not necessarily whether the phase angle can be accurately retrieved. This assumes that the planetary orbit is relatively well-constrained, consistent with the underlying assumption of this work that the planet location is known a priori.
The cloud parameters are also fixed rather than fitted during the retrievals, both because the 250--500~nm band alone provides insufficient constraints on the cloud properties and because the Hartley-Huggins ozone absorption occurs almost entirely above the cloud layer \citep{damiano_reflected_2023}. 

For the retrievals, the $\text{EXOREL}^{\mathcal{R}}$ forward model is coupled with a nested sampling algorithm, $\text{MULTINEST}$ \citep{feroz_multinest_2009, buchner_x-ray_2014}, for Bayesian parameter estimation. The retrievals are conducted with 1500 live points and a convergence tolerance of 0.5.

\begin{table}[]
    \caption{List of the free parameters used in the retrieval analysis presented in this work. All molecular abundances are retrieved as partial pressures.}
    \centering
    \begin{tabular}{clcc}
    \toprule
        \textbf{Parameter} & \textbf{Description} & \textbf{Prior Type} & \textbf{Prior Range} \\
        \midrule
        $A_g$ & Surface Albedo & linear-uniform & [0, 0.5]  \\
        $R_p$ & Planet Radius ($R_\oplus$) & linear-uniform &  [0.5, 10]\\
        $H_2S$ & Hydrogen Sulfide & log-uniform & [-7, 7] \\
        $SO_2$ & Sulfur Dioxide & log-uniform &  [-7, 7]\\
        $N_2$ & Nitrogen & log-uniform &  [-7, 7]\\
        $O_2$ & Oxygen & log-uniform & [-7, 7]\\
        $O_3$ & Ozone & log-uniform &  [-7, 7]\\
    \bottomrule
    \end{tabular}
    \label{tab:free_params}
\end{table}

\section{Ozone Detection Analysis}
\label{sec:bayes_factor}

To establish the significance of detected ozone abundances in the retrieval results across the parameter space, this work employs a statistical analysis that consists of the Anderson-Darling goodness-of-fit test \citep{stephens_edf_1974} and Bayes factor analysis \citep{thorngren_bayesian_2025, trotta_bayes_2008}. 

The Anderson-Darling (A-D) test is applied to the normalized residuals between the data and the best-fit model from the retrieval, defined as

\begin{equation}
    z_i = \frac{x_i - \hat{x}_i}{\sigma_i},
\end{equation}
where $x_i$ is the measured data value, $\hat{x}_i$ is the best-fit model value, and $\sigma_i$ is the measurement uncertainty, each evaluated at wavelength $i$. The A-D test compares the normalized residuals to a standard normal distribution ($\mu=0, \sigma^2=1$) and returns a p-value under the null hypothesis that the residuals are drawn from this distribution. The retrievals are evaluated at the $5 \%$ significance level. If null hypothesis is accepted ($p \geq 0.05$), it suggests that variation between the data and the best fit model is consistent with the photometric error of the data. If the null hypothesis is rejected ($p < 0.05$), it suggests that the best fit model cannot capture the data within the bounds of the photometric error, indicating a systematic mismatch rather than measurement noise. The A-D test therefore serves as an absolute measure of the residual consistency, indicating whether the best-fit model adequately reproduces the data within measurement uncertainty. 

The Bayes factor is instead a relative metric that is calculated as a ratio of the Bayesian evidence, $\mathcal{Z}$, of two competing models and quantifies the preference for one model over another. This work focuses on constraining the ozone abundance in modern Earth atmospheres. However, there is risk for possible degenerate solutions with compounds such as $\text{SO}_2$, which also has a strong absorption feature in the UV though it is not present in modern Earth atmospheres \citep{damiano_reflected_2023}. 

To address this degeneracy and quantify the detection significance of the ozone feature, the Bayes factor analysis compares four retrieval models, each with a different subset of molecular abundances treated as free parameters. The four models are summarized in Table~\ref{tab:bf_cases}. The baseline model includes all free parameters from Table~\ref{tab:free_params}: $\text{H}_2\text{S}$, $\text{SO}_2$, $\text{N}_2$, $\text{O}_2$, and $\text{O}_3$. The remaining three models remove either $\text{O}_3$, $\text{SO}_2$, or both as free parameters. From these four models, three Bayes factor quantities are computed:
\begin{align}
    \mathcal{B}_{\text{O}_3} &= \frac{\mathcal{Z}(\text{baseline})}{\mathcal{Z}(\text{no } \text{O}_3)}, \\
    \mathcal{B}_{\text{SO}_2} &= \frac{\mathcal{Z}(\text{baseline})}{\mathcal{Z}(\text{no } \text{SO}_2)}, \\
    \mathcal{B}_{\text{O}_3 \vee \text{SO}_2} &= \frac{\mathcal{Z}(\text{baseline})}{\mathcal{Z}(\text{no } \text{O}_3 \text{ or } \text{SO}_2)},
\end{align}
which respectively quantify the preference for including $\text{O}_3$, $\text{SO}_2$, or either molecule in the model. 
An absolute test of residual normality is a necessary complement to the Bayes factor classification, since model preference alone does not say anything about whether either model adequately explains the data \citep{welbanks_challenges_2026}. This is particularly important in this study where incomplete post-processing can result in residual systematics in the data that cannot be reproduced by the self-consistent forward model in $\text{EXOREL}^{\mathcal{R}}$, leading to a poorly converged retrieval. 

\begin{table}[h]
\caption{Molecular abundances designated as free parameters in each model}
\centering
\label{tab:bf_cases}
\begin{tabular}{lccccc}
\midrule
\textbf{Model} & \textbf{SO$_2$} & \textbf{H$_2$S} & \textbf{N$_2$} & \textbf{O$_2$} & \textbf{O$_3$} \\
\midrule
Baseline & $\checkmark$ & $\checkmark$ & $\checkmark$ & $\checkmark$ & $\checkmark$ \\
No O$_3$ & $\checkmark$ & $\checkmark$ & $\checkmark$ & $\checkmark$ &  \\
No SO$_2$ &  & $\checkmark$ & $\checkmark$ & $\checkmark$ & $\checkmark$ \\
No O$_3$ or SO$_2$ &  & $\checkmark$ & $\checkmark$ & $\checkmark$ &  \\
\midrule
\end{tabular}
\end{table}

Using the A-D test and the three Bayes factors, each retrieval outcome is classified into one of three categories: \textit{ozone detection}, \textit{ambiguous detection}, and \textit{non-detection}. 
An \textit{ozone detection} requires passing the A-D test and $\mathcal{B}_{\text{O}_3} > 3.2$ and $\mathcal{B}_{\text{SO}_2} < 3.2$, indicating that the data prefer a model with ozone but not one with $\text{SO}_2$; detections are further sub-classified as weak ($3.2 < \mathcal{B}_{\text{O}_3} \leq 10$), moderate ($10 < \mathcal{B}_{\text{O}_3} \leq 31.6$), or strong ($\mathcal{B}_{\text{O}_3} > 31.6$) following Table 1 of \cite{thorngren_bayesian_2025}. 
An \textit{ambiguous detection} occurs when the A-D test is passed and  $\mathcal{B}_{\text{O}_3} < 3.2$ but $\mathcal{B}_{\text{O}_3 \vee \text{SO}_2} > 3.2$, indicating that UV absorption is significantly preferred but cannot be unambiguously attributed to $\text{O}_3$ over $\text{SO}_2$. 
\textit{Non-detection} is assigned in two scenarios: when the  A-D test rejects the null hypothesis, indicating that the best-fit model does not adequately describe the data and Bayes factor classification is not meaningful; or, when the A-D test is passed but all three Bayes factors are less than 3.2, indicating no significant preference for UV absorption of either species. The detection classifications are summarized in Table~\ref{tab:det_classification}.

\begin{table}[h]
\caption{Ozone detection classifications. For ozone and ambiguous detections, an accepted null hypothesis from the A-D test is a pre-requisite for detection. Non-detections can either result from a rejected A-D test or an accepted A-D test where the Bayes factor reveal no model preference.}
\centering
\label{tab:det_classification}
\begin{tabular}{lccccc}
\midrule
\textbf{Classification} & A-D Test & $\mathcal{B}_{\text{O}_3}$ & $\mathcal{B}_{\text{SO}_2}$ & $\mathcal{B}_{\text{O}_3 \vee \text{SO}_2}$ \\
\midrule
\textbf{Ozone Detection} & Accept & $ > 3.2 $ &  $ < 3.2 $ & -- \\
Strong & Accept & $ > 31.6 $ & $ < 3.2 $ & -- \\
Moderate & Accept & $(10, 31.6] $ & $ < 3.2 $ & -- \\
Weak & Accept & $ (3.2, 10] $ & $ < 3.2 $ & -- \\
\midrule
\textbf{Ambiguous Detection} & Accept & $ < 3.2 $ & $ < 3.2 $ & $ > 3.2 $\\
\midrule
\textbf{Nondetection} & Reject & $ -- $ & $ -- $ & $ --$ \\
& Accept & $ < 3.2 $ & $ < 3.2 $ & $ < 3.2 $\\
\midrule
\end{tabular}
\end{table}

\section{Results}
\label{sec:results}

\begin{figure}[t!]
    \centering
    \includegraphics[width=\linewidth]{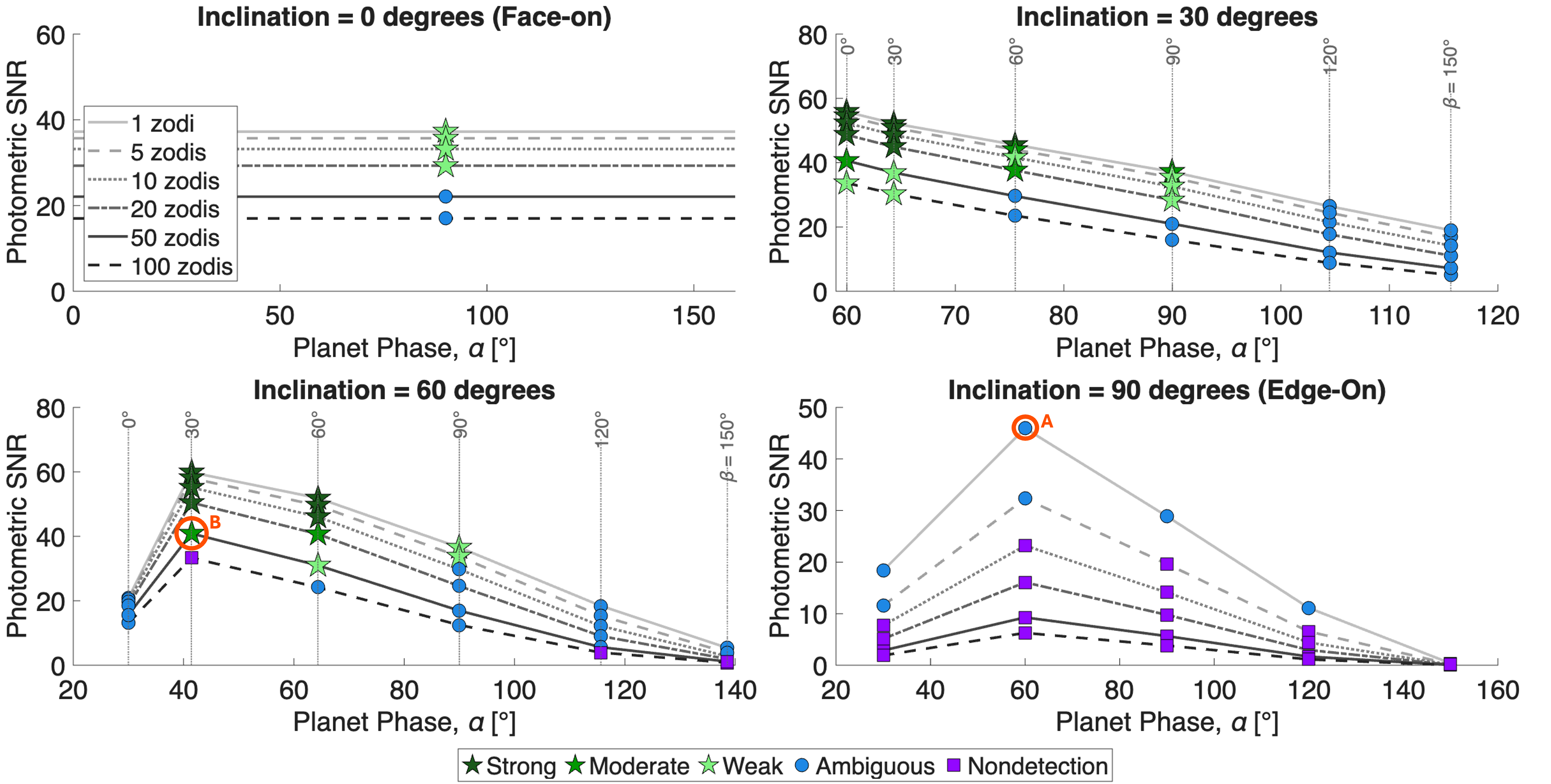}
    \caption{The Bayes factor analysis results overlaid on plots of the photometric SNR for each scenario as a function of planet phase, system inclination and exozodi density for a 3-day observation. The two annotated points, \emph{A} and \emph{B} are outlier cases that are discussed further in Section~\ref{sec:discussion_exozodi}. 
    (\emph{Top Left}) Face-on cases. Since the planet phase is $90^\circ$ regardless of the planet's longitude when the system is observed face-on, only the exozodi density is varied for these cases. (\emph{Top Right}) $30^\circ$ inclined cases. The vertical dashed lines designate the planet longitude that corresponds to each planet phase. (\emph{Bottom Left}) $60^\circ$ inclined cases. The vertical dashed lines designate the planet longitude that corresponds to each planet phase. (\emph{Bottom Right}) Edge on cases. Cases where the planet is at $0^\circ$ phase were not included because the planet is obscured by the host star when observing the system edge-on.}
    \label{fig:results_phase}
\end{figure}
\begin{figure}[t!]
    \centering
    \includegraphics[width=\linewidth]{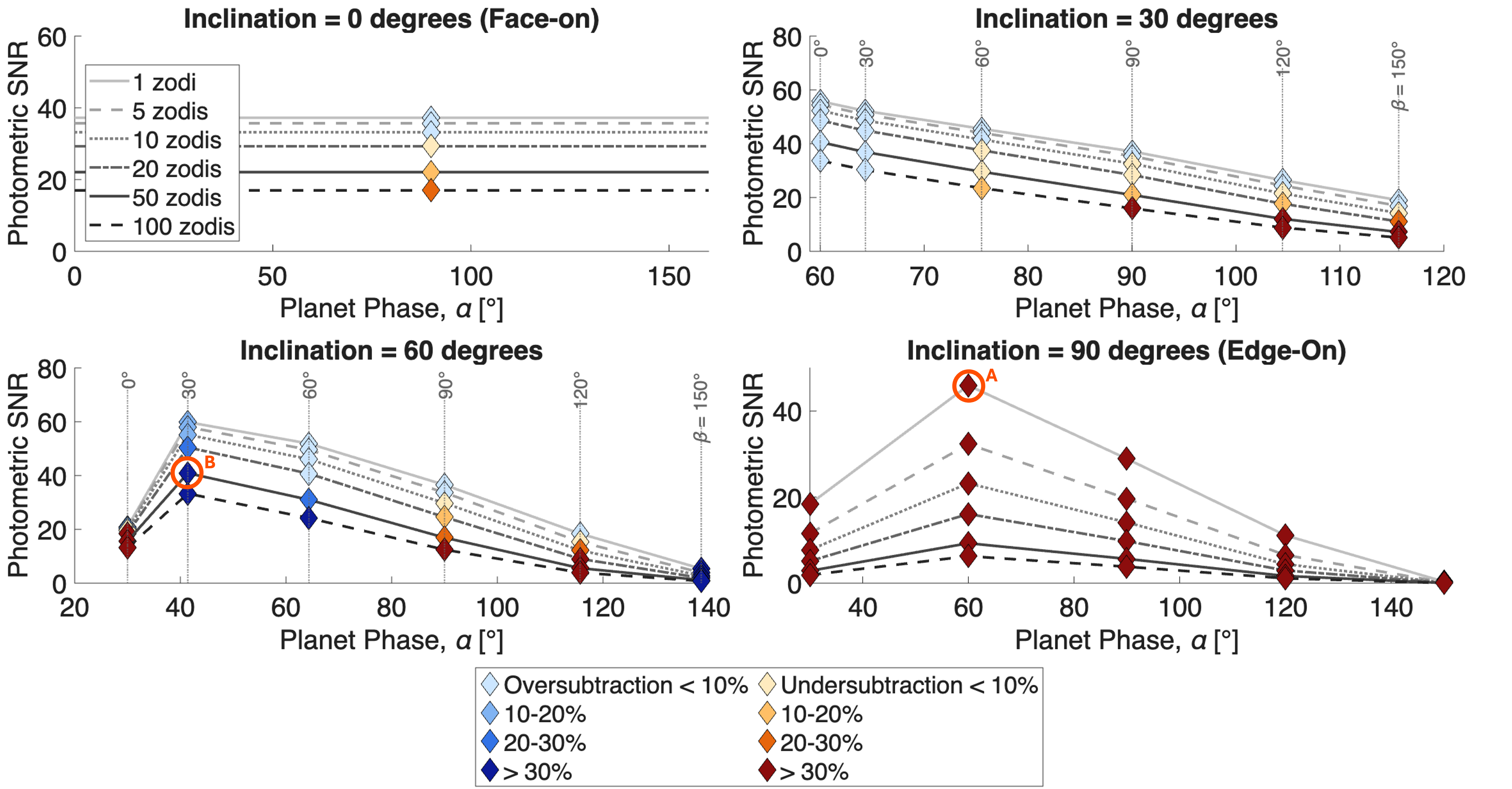}
    \caption{Similar to Figure~\ref{fig:results_phase} but with the mean bias error induced by the post-processing pipeline overlaid on plots of the photometric SNR for each scenario as a function of planet phase, system inclination and exozodi density for a 3-day observation.}
    \label{fig:results_phase_bias}
\end{figure}

To answer whether ozone can be reliably identified across a broad range of realistic observing conditions, the ozone detection analysis is applied to each scenario after performing retrievals with the models from Table~\ref{tab:bf_cases}.
Figure~\ref{fig:results_phase} shows the ozone detection  classifications 
overlaid on the photometric SNR in the 335-385 nm band, plotted as a function of planet phase, exozodi density and system inclination. Figure~\ref{fig:results_phase_bias} shows the 
mean bias induced by imperfect post-processing overlaid in place of the detection classifications.  
The photometric SNR represents the idealized, background-limited signal strength assuming perfect background subtraction and serves as a reference for interpreting the classifications across the parameter space, while the mean bias provides diagnostic context for how post-processing effectiveness modulates those results.

Although planet longitude is the parameter that is varied for the simulations, the planet phase angle sets the planet-star flux ratio and is therefore the physically meaningful quantity for interpreting the retrieval results.
Since the planet phase is a function of the longitude and inclination, uniformly sampled longitudes do not map to the same phase angles across inclinations. For reference, the corresponding planet longitude for each phase angle is shown at the top of the $30\degree$ and $60\degree$ inclination plots in Figure~\ref{fig:results_phase} and \ref{fig:results_phase_bias}. For face-on cases, the planet phase is always $90\degree$ regardless of planet longitude. For edge-on cases, the planet longitude and planet phase are equivalent.

Overall, the Bayes factor results in Figure~\ref{fig:results_phase} 
identify three broad detection regimes. For face-on and $30\degree$ inclined systems, at least weak ozone detections are achievable up to and including 20 zodis for phase angles $\leq 90\degree$, beyond which detections become ambiguous. For $60\degree$-inclined systems, at least weak detections persist up to and including 50 zodis at favorable phase angles where the planet remains outside of the starshade IWA ($30\degree < \alpha < 90\degree$), and again decrease with phase angle. Only the 100 zodi cases as unfavorable phase angles result in non-detections. Edge-on systems are the most challenging, with ambiguous detections for only the lowest zodis up to $120\degree$ and decreasing to non-detections at high zodis. 
It is worth noting that ambiguous detections, while they do not uniquely attribute UV absorption to $\mathrm{O}_3$ over $\mathrm{SO}_2$, nonetheless indicate a statistically significant preference for a UV-absorbing species, as defined in Section~\ref{sec:bayes_factor}. Such cases clearly identify a deep UV absorption feature that warrants further study, and are therefore distinct from non-detections in their scientific utility.

\begin{figure}
    \centering\includegraphics[width=0.95\linewidth]{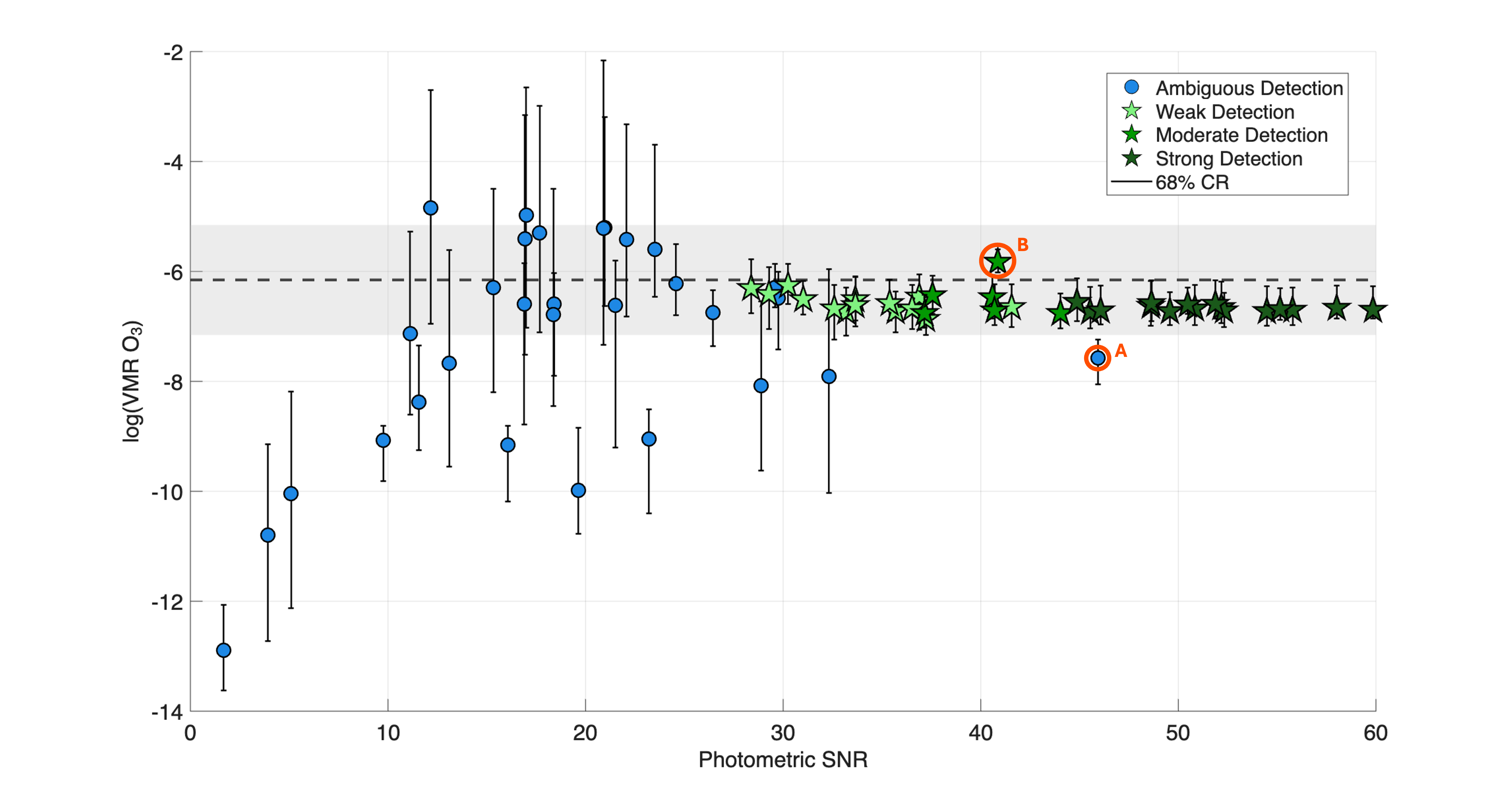}
    \caption{Photometric SNR versus the median retrieved ozone abundance. The grey shaded region designates one order of magnitude around the fiducial ozone abundance ($7 \times 10^{-7}$ VMR), shown by the dotted black line. Each case is shown with a marker at the median value and its corresponding $68\%$ credible region. The markers and their colors match those in Figure~\ref{fig:results_phase}. The two annotated points, \emph{A} and \emph{B} are outlier cases discussed in Section~\ref{sec:discussion_exozodi}. 
    }
    \label{fig:ozone_abundance}
\end{figure}

While the detection classification results assess the likelihood of the presence of ozone and its detectability across the parameter space, the median ozone abundance and its corresponding $68\%$ credible region for each scenario quantifies the accuracy and precision of the abundance constraint. Figure~\ref{fig:ozone_abundance} shows the photometric SNR versus the median retrieved ozone abundance and the $68\%$ credible region across all cases. The median value of retrieved $\mathrm{O}_3$ falls within one order of magnitude of the true input abundance in the majority cases, including many classified as ambiguous detections. This indicates the posteriors are correctly centered even where the detection significance is low. At low photometric SNR, the posteriors broaden substantially, with the $68\%$ credible region spanning several orders of magnitude in VMR $\mathrm{O}_3$. The ambiguous detection classifications in these cases reflect poor precision rather than a biased central value.

\begin{figure}
    \centering
    \includegraphics[width=0.95\linewidth]{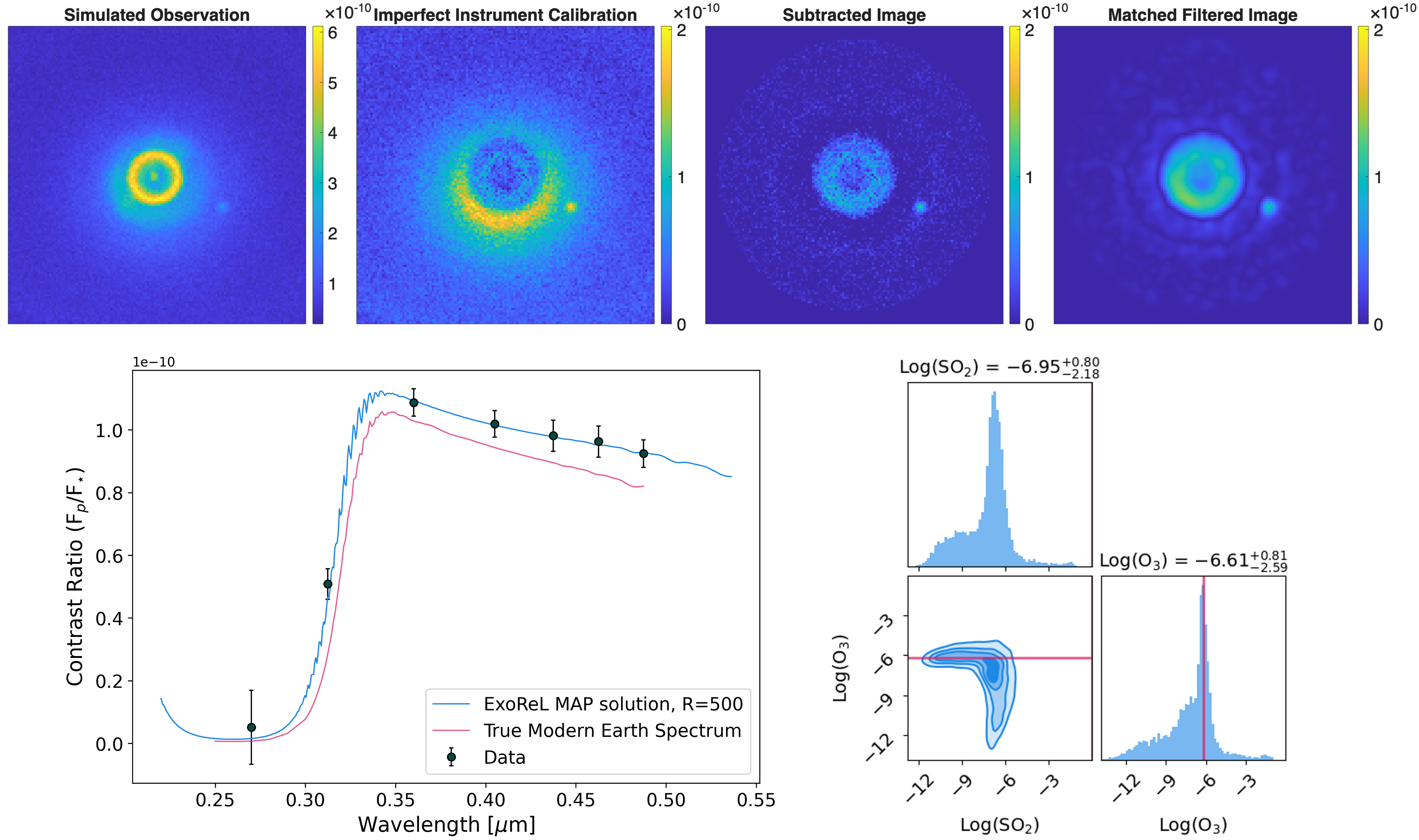}
    \caption{An ambiguous detection for a $30\degree$ inclined system with a planet phase of $\alpha = 104.5\degree$ (planet longitude $\beta = 120\degree$) and 10 zodis. The corresponding photometric SNR for this case is 21.5. (\emph{Top}) The simulated observation and output of each step of the post-processing pipeline (starshade systematic calibration, exozodi subtraction, and matched filtering)  shown for the 335-385nm band in units of contrast. (\emph{Bottom left}) The input spectral data, taken from the peak of the matched filtered image for each band, shown with the Maximum A Posteriori (MAP) solution from the retrieval as well as the true modern Earth spectrum that was used as the input for the observation simulation. (\emph{Bottom Right}) The retrieved marginal distributions for $\mathrm{SO}_2$ and $\mathrm{O}_3$ as well as their joint distribution. The values are displayed as the log of the volume mixing ratios. The magenta line designates the true $\mathrm{SO}_2$ abundance. The full posterior plots for all free parameters are included in \nameref{sec:appendix_a}.}
    \label{fig:panel_ambig}
\end{figure}

\begin{figure}
    \centering
    \includegraphics[width=0.95\linewidth]{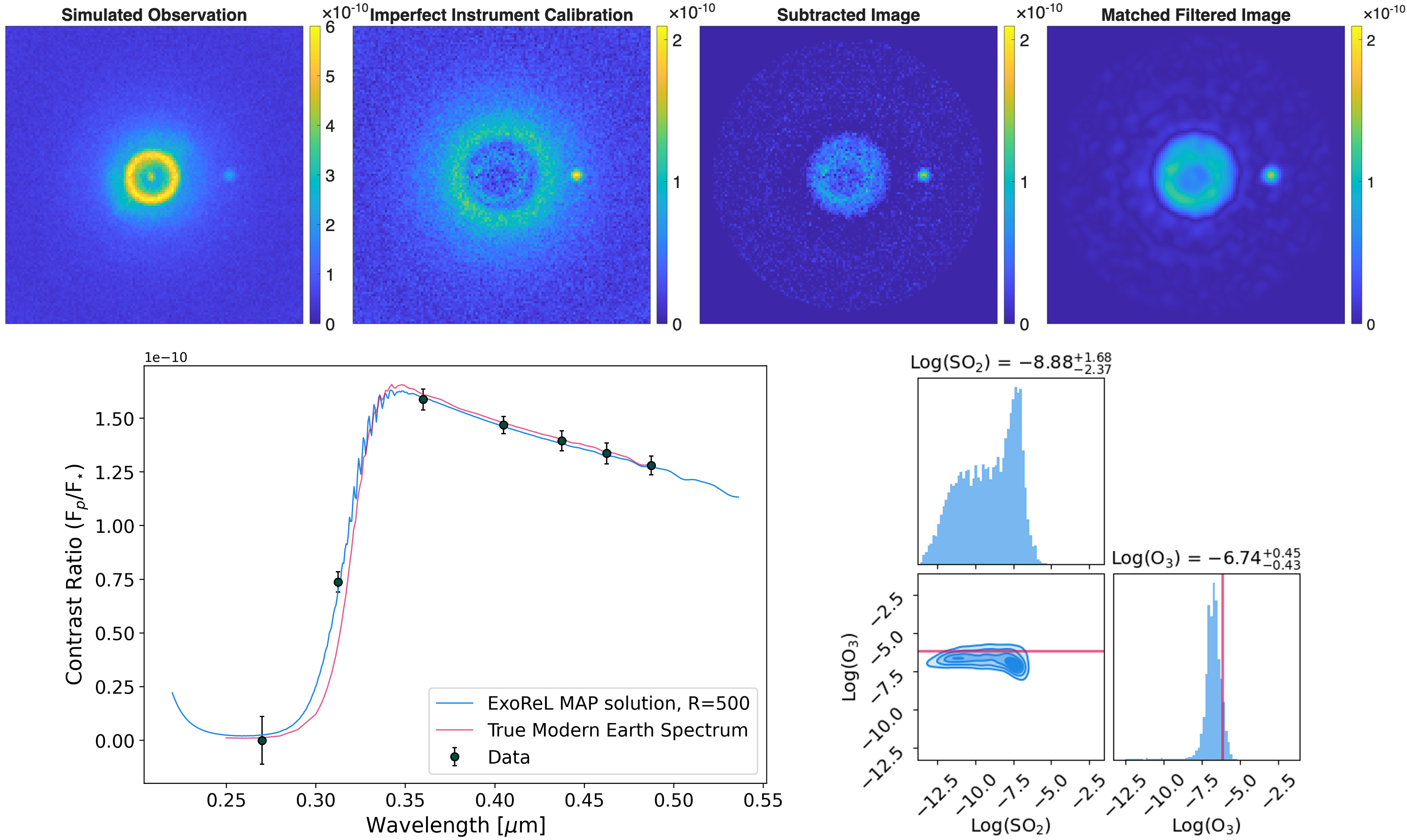}
    \caption{Similar to Figure~\ref{fig:panel_ambig} but for a weak ozone detection for a face-on system ($0\degree$ inclination) with the planet at quadrature ($\alpha = \beta = 90\degree$) and 10 zodis, and a corresponding photometric SNR of 33.2. }
    \label{fig:panel_substantial}
\end{figure}

\begin{figure}
    \centering
    \includegraphics[width=0.95\linewidth]{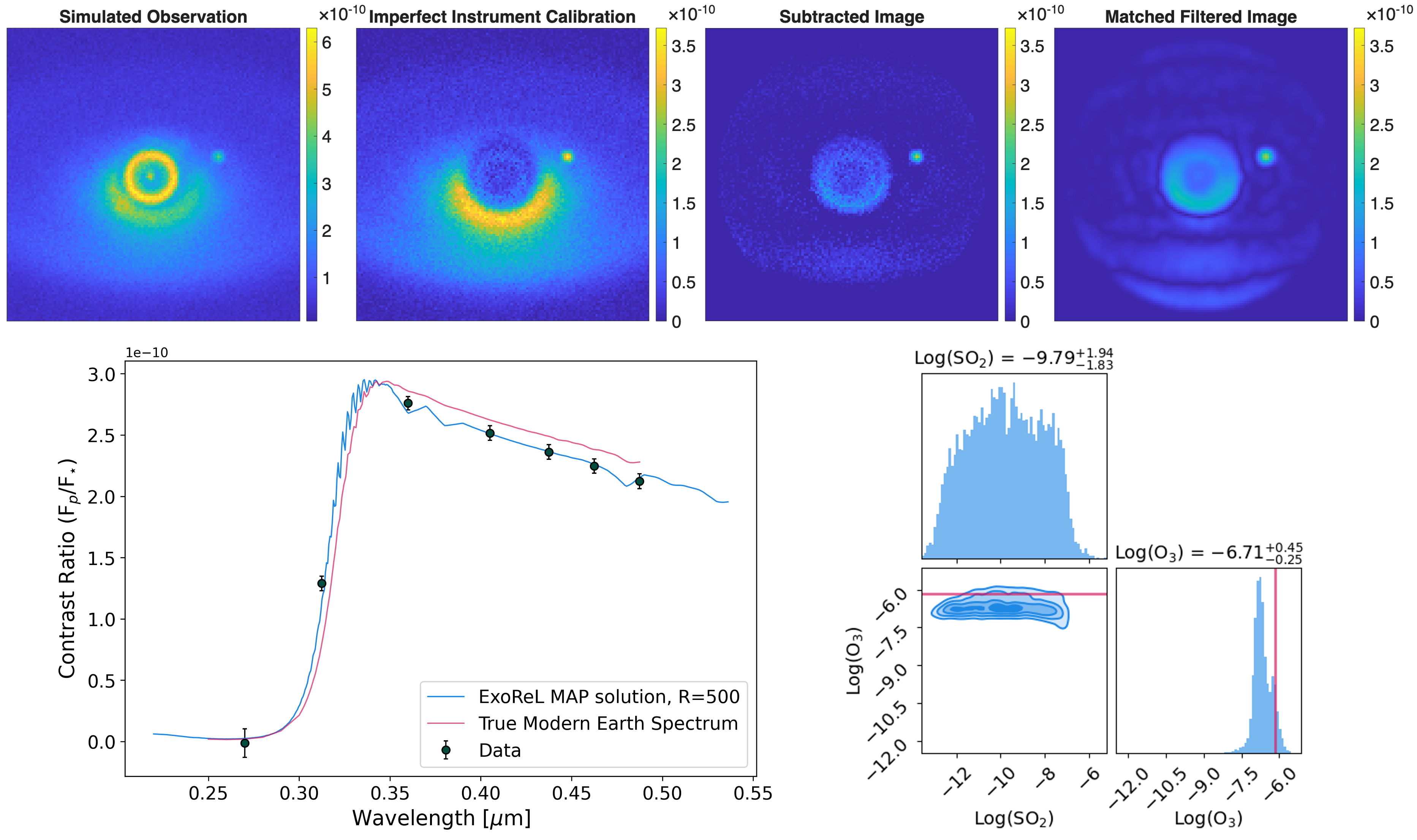}
    \caption{Similar to Figure~\ref{fig:panel_ambig} but for a strong zone detection for a $60\degree$ inclined system with a planet phase of $\alpha = 64.4\degree$ (longitude $\beta = 60\degree$) and 10 zodis, and a corresponding photometric SNR of 46.1.The background region around the planet has been slightly oversubtracted resulting in a measured spectrum that is slightly below the true input spectrum.}
    \label{fig:panel_vstrong}
\end{figure}

To provide more qualitative insight into how trends in the Bayes factor, mean bias, and median retrieved ozone abundance appear in individual retrievals, Figures~\ref {fig:panel_ambig}, \ref{fig:panel_substantial}, \ref{fig:panel_vstrong} illustrate the full simulation and retrieval pipeline outlined in Figure~\ref{fig:paper_overview} for three representative cases with different detection significances. For example, Figure~\ref{fig:panel_ambig} shows an ambiguous ozone detection for a $30\degree$ inclined system with a planet at an unfavorable phase ($\alpha = 104.5\degree$) and 10 zodis, corresponding to a photometric SNR of 21.5. The top row of panels traces the post-processing pipeline from the simulated observation through to the final matched filtered output that is used to construct the spectral data. The simulated observation shows how the planet is embedded in the forward scattering region of the exozodi disk, though because the inclination is still low, this forward scattering region is not as bright as it is for a more highly inclined system. In the subtracted image, the exozodi has not been cleanly subtracted, and there is a bright ring of residual systematics in the area around the planet. This results in a positive bias (undersubtraction of the background) in the extracted spectrum when compared to the true modern Earth spectrum that was used as an input to the simulation, as seen in the bottom left of the figure. 
The maximum-a-posteriori (i.e. best fit) solution from the ExoReL retrieval is well-fitted to the data, however, the posterior distributions highlight the ambiguity of this detection. Although the data clearly prefer the inclusion of a UV-absorbing species, the joint posterior shows substantial overlap between $\mathrm{SO}_2$ and $\mathrm{O}_3$, indicating that the retrieval does not explicitly prefer ozone, hence the ambiguous detection.

Figures~\ref{fig:panel_substantial} and \ref{fig:panel_vstrong} follow the same layout but for a weak and strong ozone detection, respectively. Across the three cases, the joint $\mathrm{SO}_2/\mathrm{O}_3$ posterior tightens, the $\mathrm{O}_3$ marginal becomes more sharply peaked, and the discrimination between $\mathrm{O}_3$ and $\mathrm{SO}_2$ improves as the SNR increases.

\section{Discussion}
\label{sec:discussion}

The results in Section~\ref{sec:results} establish where ozone can be reliably identified across the parameter space consisting of phase angle, inclination and exozodi density for the given observatory architecture. The following section examines these trends in detail, discussing the SNR requirements for UV ozone detection, the impact of exozodi systematics, the robustness of the retrieved ozone abundance, and the broader implications for UV starshade science returns.

\subsection{SNR Requirements for UV Ozone Detection}
\label{sec:discussion_snr}

The Bayes factor results shown in Figure \ref{fig:ozone_abundance} establish a photometric SNR threshold of $\sim 30$ to achieve at least a weak ozone detection with UV-only, $R=7$ spectroscopy. 

This threshold reflects the limited information content of the observing regime; at low spectral resolution with only UV wavelength coverage (i.e. no additional information from the visible or NIR bands), a high SNR is required to confidently attribute the absorption feature to $\mathrm{O}_3$ over $\mathrm{SO}_2$. \cite{damiano_reflected_2023} found that detection of $\mathrm{O}_3$ was possible at SNR 9--12 with a spectral resolution of $R = 7$ in the $0.25-0.4\mu\mathrm{m}$ range for Proterozoic Earth atmospheres. However, that work used UV measurements as supplements to visible and NIR measurements to assess the added value of UV coverage rather than its standalone capability. Additionally, the two studies vary in their definition of SNR. \cite{damiano_reflected_2023} fixed the SNR to 20 in the visible wavelengths and then extended the same error bar size to all other wavelengths, resulting in the lower SNR 9--12 for the UV measurements. In contrast, the error bars in this work vary by wavelength as a direct output of the starshade post-processing pipeline. 
The higher SNR threshold found in this work therefore compensates for the additional discriminating power that multi-wavelength coverage would provide, and establishes the baseline SNR requirement specific to UV-only ozone characterization with non-uniform measurement errors.

\subsection{Impact of Residual Exozodi Systematics}
\label{sec:discussion_exozodi}

While the Bayes factor results demonstrate the $\mathrm{SNR}\sim30$ threshold, it is important to understand how the residual post-processing systematics influence these results. \cite{ahmed_ultraviolet_2025} showed that the starshade systematics have minimal impact on the recovered signal when the planet is located outside of the starshade's IWA and highlighted the residual exozodiacal disk emission as the dominant source of post-processing bias. The remainder of this section therefore focuses on the residual exozodi systematics.

Both exozodi density and system inclination are the primary drivers of the exozodi subtraction fidelity. This can be seen in the mean bias results from Figure~\ref{fig:results_phase_bias}, where the mean bias grows with exozodi density and is larger on average as the inclination increases. Higher exozodi densities produce larger absolute residuals after subtraction, increasing the fractional bias on the extracted planet flux. As the inclination increases, forward scattering enhances the disk surface brightness and the disk's projected morphology becomes increasingly asymmetric, ultimately collapsing toward a sharp, knife-edge feature for edge-on systems \citep{currie_mitigating_2023}. The low-order parametric model used in this work cannot adequately capture the spatial variability of the strongly forward-scattered, spatially compact emission at high inclination. As a result, the forward scattering regions of the disk are under-
subtracted, leaving a positive residual that biases the extracted planet signal, while the backscattered regions are over-subtracted, removing a portion of the planet signal itself. Unfavorable phases correspond to the under-subtracted region of the disk and favorable phases are over-subtracted, as seen in Figure~\ref{fig:results_phase_bias}. At low exozodi densities ($< 20$ zodis), the bias is minimal ($\leq 10\%$), regardless of over- or under-subtraction. For edge-on cases, the inclination is the dominant limiting factor, with post-processing residuals overwhelming the planet signal regardless of phase angle or zodi density. 

The direction of this bias affects the Bayes factor in opposite ways. Over-subtraction deepens the apparent ozone absorption feature and inflates the recovered Bayes factor, while under-subtraction fills in the feature and biases the Bayes factor toward lower significance. This can be seen when comparing the strong detection in Figure~\ref{fig:panel_vstrong} which has been over-subtracted, with the ambiguous case in Figure~\ref{fig:panel_ambig} where the planet signal is under-subtracted. While the specific regions of the parameter space where over- versus under-subtraction occurs is dependent on the post-processing approach, these directional effects on the Bayes factor are general and apply regardless of the pipeline used.

Figure~\ref{fig:ozone_abundance} reinforces the conclusions from Section~\ref{sec:discussion_snr}. 
The SNR~$\sim 30$ threshold is clearly visible; cases above this threshold are classified as detections that increase in strength with increasing SNR and are clustered around the fiducial ozone abundance within one order of magnitude with tight $68\%$ credible regions. Cases below this SNR threshold show substantially broader posteriors consistent with ambiguous Bayes factor classifications. 

Two cases, labeled \emph{A} and \emph{B} in Figure \ref{fig:ozone_abundance}, stand out for their outlier ozone abundances.  Case \emph{A} is under-subtracted by $33\%$, resulting in a suppression of the ozone absorption feature that leads to a correspondingly underestimated ozone abundance, as well as an ambiguous detection even though it is well above SNR $>$ 30.  When the bias is removed and the exozodi is perfectly subtracted, the ozone abundance shifts from $-7.6^{+0.3}_{-0.5}$ to $-6.6^{+0.4}_{-0.5}$, and the detection significance improves to a moderate ozone detection, consistent with high SNR data with lower bias.  Case \emph{B} is instead over-subtracted by $48\%$ with an SNR of 41, resulting in an over-estimated ozone abundance with a tight credible region. When the exozodi is perfectly subtracted, case \emph{B} still recovers a moderate ozone detection but the abundance shifts from $-5.8^{+0.2}_{-0.12}$ to $-6.5^{+0.4}_{-0.4}$, consistent with the rest of the detected cases. 
These two examples show that significant ozone abundance estimation error, and in some cases, a too-weak or too-strong Bayes classification, occurs when the magnitude of the signal estimation bias is $>$ $\sim$ 30\%.

\subsection{Science Return of Low-Resolution Starshade Spectroscopy}

UV starshade spectroscopy provides a distinct and complementary constraint on atmospheric composition relative to visible and NIR observations. The Hartley-Huggins ozone band at $0.25 \mu\mathrm{m}$ is detectable at low spectral resolution down to $0.1\%$ PAL $\mathrm{O}_2$, making it accessible even for Proterozoic-like atmospheres in which the visible and NIR oxygen features become inaccessible \citep{schwieterman_exoplanet_2018, reinhard_false_2017, Lisman_surveying_2019, schwieterman_importance_2018}. The results here demonstrate that this constraint can be obtained efficiently with short UV starshade observations alone, establishing the standalone capability of UV spectroscopy as a complement to the multi-wavelength analysis from \cite{damiano_reflected_2023}. Achieving a photometric SNR of 30 with a UV starshade paired with an 6-m HWO telescope requires 61 hours of observing time for a $30\degree$ inclined system with a planet at quadrature and 10 zodis. If the planet has a more favorable phase angle, such as a $60\degree$ inclined system with a planet observed at a $41\degree$ phase angle and 20 zodis, the same SNR threshold can be achieved with 25 hours of integration time. Even for ambiguous detections that cannot distinguish $\mathrm{O}_3$ from $\mathrm{SO}_2$ with UV observations alone, these short starshade observations can be used as motivation for further characterization to search for the $\mathrm{O}_2$ A-band at 0.76 $\mu\mathrm{m}$ or water vapor bands at 0.94 $\mu \mathrm{m}$. Substantial atmospheric water vapor would rule out $\mathrm{SO}_2$ as the primary UV absorber, as sustained atmospheric $\mathrm{SO}_2$ is incompatible with significant surface liquid water \citep{loftus_sulfate_2019}.

The results from Section~\ref{sec:discussion_snr}, \ref{sec:discussion_exozodi} 
together establish the science return of UV starshade observations. The starshade architecture is capable of reaching the photometric SNR~$\sim 30$ threshold required for robust ozone detection within 3-day integrations for face-on through $60\degree$-inclined systems at favorable phase angles. The exozodi systematics results then establish that detections at this SNR are robust to subtraction bias within approximately $30\%$, which is achievable for the majority of favorable-phase cases even with a simple and conservative post-processing pipeline such as the one used in this work.

\section{Conclusions}
\label{sec:conclusion}

This work investigates the science return of low-resolution UV spectroscopy with 3-day starshade observations, assessing how phase angle, system inclination and residual systematic biases from exozodiacal dust affect ozone detectability in modern Earth atmospheres. The analysis uses a novel end-to-end framework that simulates the starshade observations with all relevant instrument and astrophysical systematics, applies a post-processing routine, and finally conducts Bayesian spectral retrieval coupled with the Anderson-Darling test and Bayes factor analysis to quantify ozone detection significance. The approach used in this work addresses a gap in prior retrieval studies by allowing systematic residuals to propagate into the simulated spectral data rather than assuming idealized noise properties.

The results establish two thresholds for UV-only, $R=7$ reflected-light spectroscopy. First, a photometric SNR~$\sim 30$ is required for robust ozone detection. Second, within this SNR regime, detections are robust to exozodi subtraction bias within $30\%$, beyond which systematic distortion of the extracted spectrum can lead to biased retrieval results. Below the SNR threshold, cases are photometrically limited and improved post-processing does not recover an un-ambiguous detection of ozone. Even in cases where ozone cannot be disambiguated from degenerate solutions including sulfur dioxide, the median retrieved ozone abundance remains accurate to within one order of magnitude of the fiducial value, indicating that ambiguous cases would likely be confirmed with follow-up observations with longer integration times or when the planet is at a more favorable phase.

More broadly, these results highlight the power and operational role of UV starshade observations for future missions such as the Habitable Worlds Observatory. The combination of the starshade's deep contrast, high throughput, small inner working angle, and broad UV bandpass enables characterization 
of ozone-bearing atmospheres with only a few days of integration time, providing constraints on a key biosignature that complements the broader atmospheric characterization performed at visible and NIR wavelengths. In this role, UV starshade observations could also serve as an useful ``triage" step for previously detected planet candidates, identifying those with oxygen-bearing chemistry that warrant more extensive follow-up needed for full atmospheric characterization. This holds even for ambiguous ozone detection, where the planet can be identified as a target for visible and NIR follow-up observations to conclusively rule out $\mathrm{SO}_2$ via strong water vapor detection.

By bridging high-fidelity imaging simulations and Bayesian spectral retrievals, the framework developed here provides a foundation for assessing the science return of future direct imaging observations. Future work will employ this approach for broader atmospheric types, such as Proterozoic Earths, where reduced $\mathrm{O}_2$ abundances make the UV ozone feature an even more critical biosignature constraint.

\section{Acknowledgments}

The authors would like to thank Amber Young, Sara Seager, Jessie Christiansen and Renyu Hu for the helpful discussions. 
Computations were performed using the computer clusters and data storage resources of the University of California, Riverside HPCC, which were funded by grants from NSF (MRI-2215705, MRI-1429826) and NIH (1S10OD016290-01A1).
Additionally, computing for this project was performed on the Sherlock cluster at Stanford University. The authors would like to thank Stanford University and the Stanford Research Computing Center for providing computational resources and support that contributed to these research results.
This work was partially supported by the National Science Foundation Graduate Research Fellowship (Grant No. DGE-1656518). EWS acknowledges support from the Interdisciplinary Consortium for Astrobiology Research (ICAR) Program through the Alternative Earths team, with funding issued under grant No. 80NSSC21K0594, and the Virtual Planetary Laboratory, with funding issued under grant No. 80NSSC23K1398.
Part of this research was carried out at the Jet Propulsion Laboratory, California Institute of Technology, under a contract with the National Aeronautics and Space Administration (80NM0018D0004).

\bibliography{main}{}
\bibliographystyle{aasjournalv7}

\section*{Appendix A}
\label{sec:appendix_a}

While the results and discussion in Sections~\ref{sec:results} and \ref{sec:discussion} focus on ozone abundance and the ability to discriminate it from sulfur dioxide, the retrieval results include all parameters listed in Table~\ref{tab:free_params} and their posterior plots, or corner plots, are included here.

\begin{figure}[h!]
    \centering
    \includegraphics[width=\linewidth]{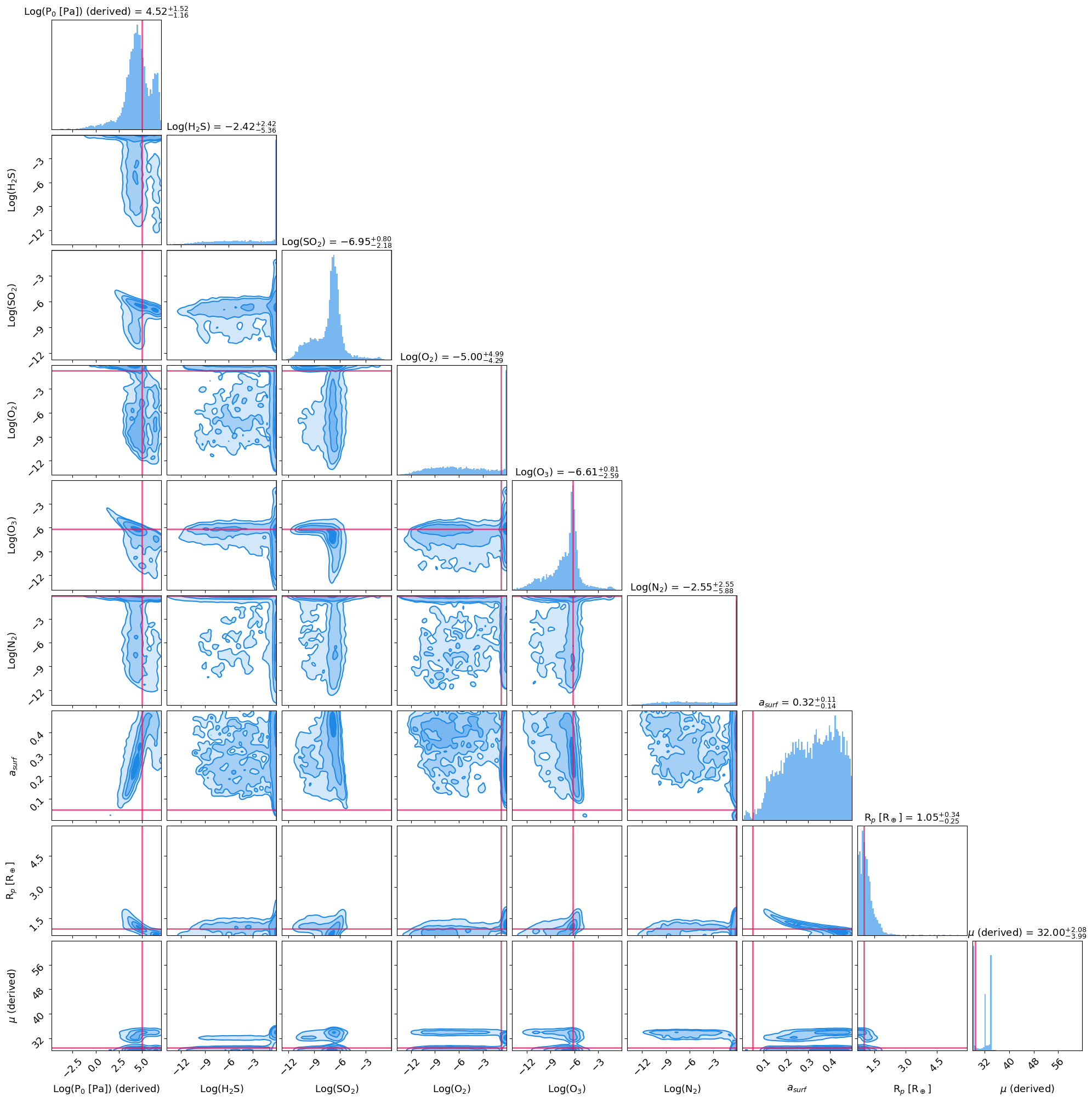}
    \caption{The full posterior distributions of all parameters included in the retrieval process for the $30\degree$-inclined system with 10 zodis and a planet with phase $104.5\degree$ shown in Figure~\ref{fig:panel_ambig}. The magenta lines show the true values.}
    \label{fig:fullpost_ambig}
\end{figure}

\begin{figure}[h!]
    \centering
    \includegraphics[width=\linewidth]{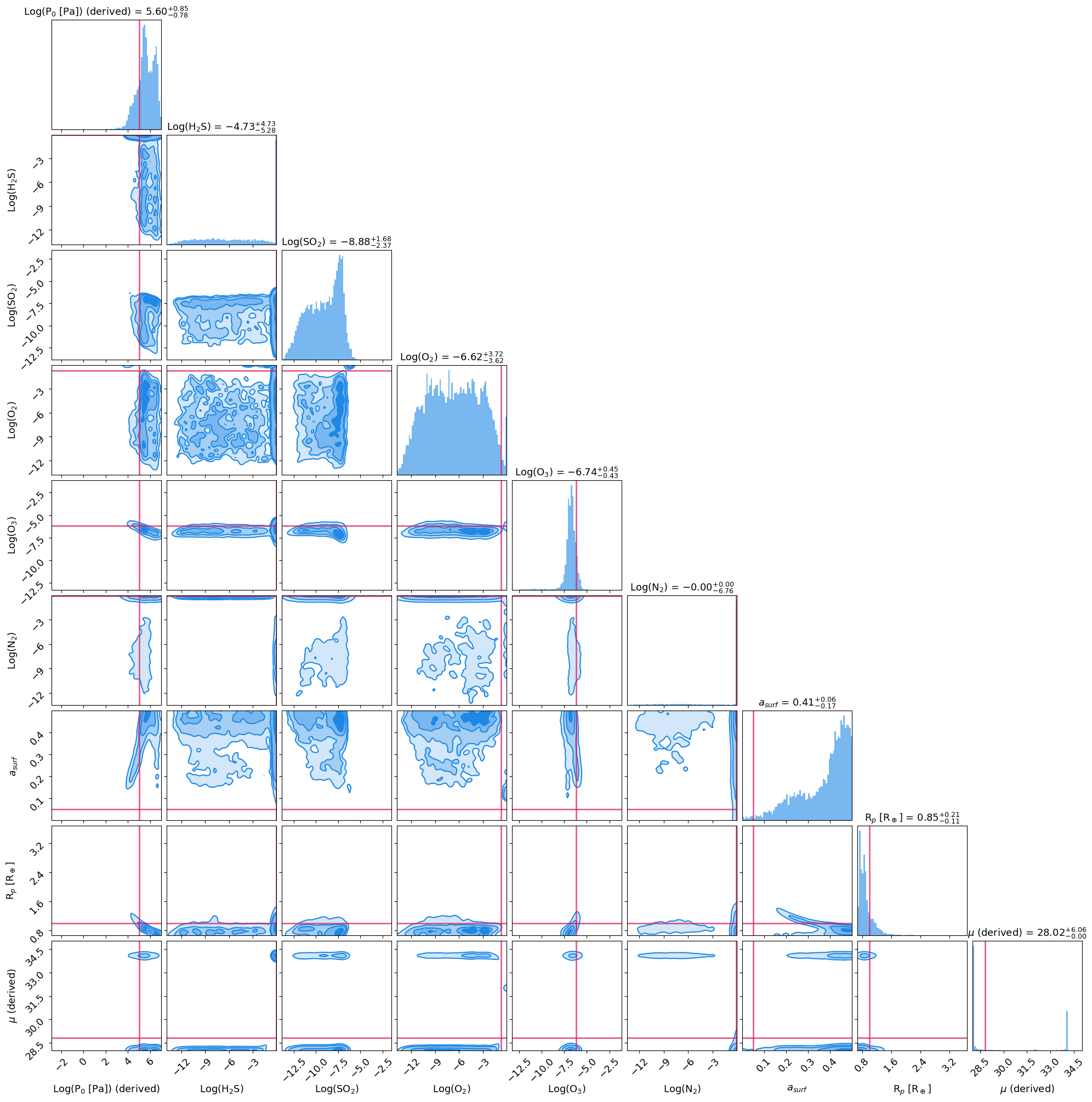}
    \caption{The full posterior distributions of all parameters included in the retrieval process for the face-on system with 10 zodis shown in Figure~\ref{fig:panel_substantial}. The magenta lines show the true values.}
    \label{fig:fullpost_weak}
\end{figure}

\begin{figure}[h!]
    \centering
    \includegraphics[width=\linewidth]{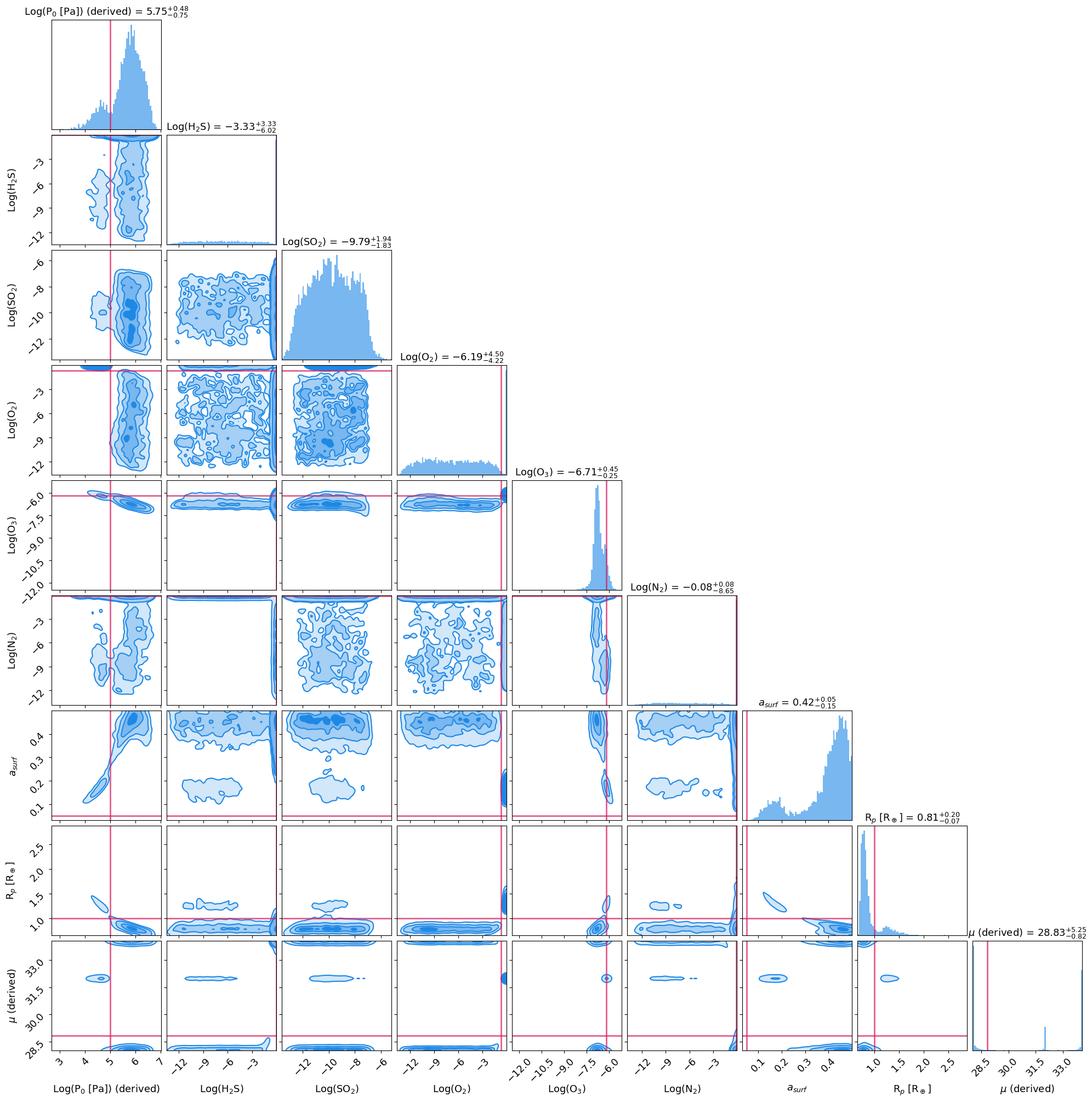}
    \caption{The full posterior distributions of all parameters included in the retrieval process for the $60\degree$-inclined system with 10 zodis and a planet with phase $104.5\degree$ shown in Figure~\ref{fig:panel_vstrong}. The magenta lines show the true values.}
    \label{fig:fullpost_strong}
\end{figure}

\end{document}